\documentclass[12pt,a4paper]{article}
\usepackage{endnotes}
\let\footnote=\endnote

\usepackage{natbib}
\bibpunct[, ]{(}{)}{,}{a}{}{,}%
\usepackage{color,soul}

\usepackage{booktabs, threeparttable, longtable, lscape, rotating}

\newcommand{\resethlcolor}{\sethlcolor{yellow}}\resethlcolor
\definecolor{lightblue}{rgb}{.90,.95,1}
\definecolor{bluegreen}{rgb}{0,1,0.5}

\newcommand{\wang}[1]{\color{green}{[#1 (Wang)]}\color{black}}

\newcommand{\al}[1]{\color{black}{#1}\color{black}}

\fontsize {12 pt}{40 cm}\selectfont

\renewcommand\b[1]{\mbox{\bf #1}}

\usepackage{amsfonts}
\usepackage{amsmath}
\usepackage{amssymb}
\usepackage{natbib}
\usepackage{graphicx}
\usepackage{graphics}

\usepackage{booktabs}
\usepackage{threeparttable}
\usepackage{longtable,lscape}
\newtheorem{theorem}{Theorem}

\newtheorem{corollary}{Corollary}
\newtheorem{proposition}{Proposition}
\begin{document}

\bibliographystyle{chicago}
\title{Efficient Importance Sampling for Rare Event Simulation with Applications}
\author{Cheng-Der Fuh\thanks{Chair Professor at the Graduate Institute of Statistics, National Central University, No. 300, Jhongda
Rd., Jhongli City, Taoyuan County 32001, Taiwan (R.O.C.).
Email:cdfuh@ncu.edu.tw},~~~Huei-Wen Teng\thanks{Assistant Professor at the
Graduate Institute of Statistics, National Central University, No. 300, Jhongda
Rd., Jhongli City, Taoyuan County 32001, Taiwan (R.O.C.). Email:wenteng@ncu.edu.tw}
~~~and~~~Ren-Her Wang\thanks{Assistant Professor at the
Department of Banking and Finance, Tamkang University, New Taipei City, 25137 Taiwan (R.O.C.).
Email:138230@mail.tku.edu.tw}}
\date{\today}
\maketitle
\bigskip

\begin{abstract}
Importance sampling has been known as a powerful tool to reduce the
variance of Monte Carlo estimator for rare event simulation. Based
on the criterion of minimizing the variance of Monte Carlo
estimator within a parametric family, we propose a general account for finding the
optimal tilting measure. To this end, when the moment generating function of the underlying distribution
exists, we obtain a simple and explicit expression of the optimal alternative
distribution. The proposed algorithm is quite general to cover many interesting
examples, such as normal distribution, noncentral $\chi^2$ distribution, and compound Poisson processes.
To illustrate the broad applicability of our method, we study value-at-risk
(VaR) computation in financial risk management and bootstrap
confidence regions in statistical inferences. 
\end{abstract}


\vspace{0.3in} \noindent {\it Some key words:} Bootstrap; confidence
region; exponential tilting; moderate deviation; VaR. \\

\newpage
\section{Introduction}
\def\theequation{1.\arabic{equation}}
\setcounter{equation}{0}

This paper considers the problem of estimating small probabilities
by Monte Carlo simulations. That is, we estimate $z = P(A)$ when $z$
is small, say of the order $10^{-2}$ or $10^{-3}$ or so; i.e., $A$
is a moderate deviation rare event.
Such problems appear in the construction of confidence regions for
asymptotically normal statistics; cf. Beran (1987), Beran and Millar
(1986), Hall (1987, 1992), Fuh and Hu (2004), and in the computation
of value-at-risk (VaR) in risk management; cf. Jorion (2001), Duffie
and Singleton (2003), Glasserman et al. (2000, 2002), Fuh et al.
(2011). It is well known that importance sampling, where one uses
observations from an alternative distribution $Q$ to estimate the
target distribution $P$, is a powerful tool in efficient simulation
of events with small probabilities. Some general references
are in Heidelberger (1995), Liu (2001), and Asmussen and Glynn (2007).

A useful tool in importance sampling for rare event simulation is
exponential tilting, cf. Siegmund (1976), and Bucklew (2004) and references therein.
The above mentioned algorithm is more efficient for large deviation rare event, i.e., $z$ is of
the order $10^{-5}$ or less. Examples of such events occur in telecommunications
($z=$ bit-loss rate, probability of buffer overflow) and reliability
($z=$ the probability of failure before time $t$). To be more
precise, when a sequence of random vectors $\{X_n\}$ converge to a
constant vector $\mu$, for any event $A$ not containing $\mu$, the
probability $P\{X_n\in A\}$ usually decays exponentially fast as
$n\rightarrow\infty$. Efficient Monte Carlo simulation of such
events has been obtained by Sadowsky and Bucklew (1990) based on the
large deviations theory given by Ney (1983).

For moderate deviation rare event simulations, efficient importance
sampling has been studied by \cite{j}, \cite{da}, \cite{dh}, and Fuh and
Hu (2004, 2007). However, those papers concern one- and/or
multivariate-normal distributions. Extension to heavy-tailed
settings such as multivariate $t$ distribution can be found in Fuh
et al. (2011). The goal of this paper is to provide a simple general account for
exponential tilting importance sampling, which covers all previous
results and many other interesting examples.

It is worth mentioning that for events of large deviations $P\{X \in
A\}$, Sadowsky and Bucklew (1990) showed that the asymptotically
optimal alternative distribution is obtained through exponential
tilting; that is, $Q(dx)=C\exp(\theta x)P(dx)$, where $C$ is a
normalizing constant and $\theta$ determines the amount of tilting.
The optimal amount of tilting $\theta$ is such that the expectation of X under $Q$-measure
equals the dominating point located at the boundary of $A$. However,
for moderate deviation rare event, we show that typically the
tilting point of the optimal alternative distribution is in the
interior of $A$, which is different from the dominating point of
large deviations theory. Furthermore, by using the idea of conjugate measure of $Q$,
$\bar{Q}(dx)=C\exp(-\theta x)P(dx)$, the general account of our
approach characterizes the optimal tilting $\theta$, by solving the equation of the expectation of $X$
under $Q$-measure equals the conditional expectation of $X$ under $\bar{Q}$-measure given the rare event.

There are three aspects in this study. To begin with, we obtain an
explicit expression for the optimal alternative distribution
under exponential embedding family.
Second, the proposed algorithm is quite general to cover many
interesting examples, such as normal distribution, noncentral $\chi^2$ distribution, and compound Poisson processes.
Third, the derived tilting formula can be used to calculate portfolio VaR
under jump diffusion models, and to approximate bootstrap confidence regions of
parameters in regression models.

The rest of this paper is organized as follows. In Section 2, we
present a general account of importance sampling that minimizes the
variance of the Monte Carlo estimator within a parametric family, provide a recursive
algorithm for finding the optimal alternative distribution, and approximate optimal tilting probability
measure for moderate deviation events.
Section 3 presents several frequently used examples to which we can characterize the
optimal titling probability measures,
and reports the relative efficiency of the proposed
method with respect to the naive Monte Carlo through a simulation study.
In Section 4, we demonstrate the performance of the tilting formula
by investigating two examples: calculating portfolio VaR and bootstrapping confidence
regions. Concluding remarks are
given in Section 5. The proofs are deferred to the appendix.

\section{Importance Sampling}
\def\theequation{2.\arabic{equation}}
\setcounter{equation}{0}
\subsection{A general account in importance sampling}
Let $(\Omega, {\cal F}, P)$ be a given probability space, $X $ be a random variable on $\Omega$ and
$A$ be a measurable set in ${\bf R}$.
To estimate the probability of an event $\{X \in A\}$, we shall
employ the importance sampling method. That is, instead of sampling
from the target distribution $P$ of $X$ directly, we sample from an
alternative distribution $Q:=Q_\theta$. Suppose $X$ has moment generating
function $\Psi(\theta)=E[e^{\theta X}]$ under $P$ for $\theta \in {\bf R}$. Then we consider
the exponential tilting measure $Q$ of $P$, which has the form
$$\frac{dQ}{dP}=\frac{\text{e}^{\theta X}}{E[\text{e}^{\theta X}]}=\text{e}^{\theta X -\psi(\theta)},$$
where  $\psi(\theta)$ is $\log \Psi(\theta)$, the cumulant
generating function of $X$. 
The question is how to choose an
alternative distribution $Q$ so that the importance
sampling estimator has the minimum variance.

The importance sampling estimator for $p=P\{X \in A\}$ based on a
sample of size $n$ is
\begin{equation}\label{ise}
\hat{p}_n=\frac{1}{n}\sum_{i=1}^{n}\textbf{1}_{\{X_i \in A \}}\frac{d{P}}{d{Q}},
\end{equation}
where $\textbf{1}_B$ is the indicator function of an event $B$,
$X_i,~i=1,\ldots,n,$ are independent observations from $Q$, and
$d{P}/d{Q}$ is the Radon-Nikodym derivative assuming $P$ is
absolutely continuous with respect to $Q$. Set
\begin{eqnarray*}
G(\theta)=E_{Q}\bigg[\textbf{1}_{\{X \in A\}}\frac{d{P}}{d{Q}}\bigg]^2=E\bigg[\textbf{1}_{\{X \in A\}}\frac{dP}{dQ}\bigg]=
E[\textbf{1}_{\{X\in A\}}\text{e}^{-\theta X + \psi(\theta)}],
\end{eqnarray*}
where the expectation without any qualification is under the target
probability measure ${P}$ unless otherwise stated. Observing that,
since the estimator $\hat{p}_n$ is unbiased, the variance of the
importance sampling estimator is
\begin{eqnarray*}
{\rm var}_{Q}(\hat{p}_n)=n^{-1}(G(\theta)-p^2).
\end{eqnarray*}

We remark that when $A=\{\omega:X(\omega)>a\}$ for some constant
$a>0$, large deviation theory considers the following inequality
\[
G(\theta)\le e^{-\theta a + \psi(\theta)},
\]
and minimizes the above upper bound. The first-order condition (after taking logarithm) gives
\begin{eqnarray}\label{g4}
\psi'(\theta)=a.
\end{eqnarray}
Note that the approximation (\ref{g4}) is more accurate when $P\{X > a\}$ is
small, i.e., $a$ is sufficient large. In contrast to the large
deviation theory, our goal is to solve the optimization problem for $\theta$
\begin{eqnarray}\label{optimization}
\theta^{\ast} = \arg\!\min\limits_{{\theta}} G(\theta).
\end{eqnarray}
$\theta^\ast$ is hence the desired quantity for selecting the tilting measure that minimizes
the variance of the importance sampling estimator within a suitable parametric family.

To minimize $G(\theta)$, the first-order condition gives
\[
\frac{\partial G(\theta)}{\partial \theta}=E[\textbf{1}_{\{X\in A\}} e^{-\theta X+\psi(\theta)}(-X+\psi'(\theta))]=0.
\]
Dividing $\text{e}^{\psi(\theta)}$ for both sides of the above equation, we have an equivalent condition,
\begin{eqnarray}\label{g2}
E[\textbf{1}_{\{X\in A\}} e^{-\theta X}(-X+\psi'(\theta))]=0,
\end{eqnarray}
and therefore $\theta^\ast$ is the solution of
\begin{eqnarray}\label{abeq}
\psi'(\theta) = \frac{E[\textbf{1}_{\{X\in A\}}X \text{e}^{-\theta X}]}{E[\textbf{1}_{\{X\in A\}} \text{e}^{-\theta X}]}.
\end{eqnarray}

Ideally, closed-formed formulas of the right-hand-side (RHS) of (\ref{abeq}) needs to be derived via analytic procedures.
Then, standard numerical procedures (or a recursive algorithm presented in Section 2.2) can be applied to find $\theta^\ast$ satisfying (\ref{abeq}).
However, the derivation for a closed-formed formula of the RHS of (\ref{abeq}) may be tedious and complicated in general cases.
In this respect, the tilting measure given in (\ref{g4}) in the large deviation theory seems to be preferred because of its simplicity. To prove that
the RHS of (\ref{abeq}) can be represented in a simple formula, we consider the conjugate measure
$\bar{Q}:=\bar{Q}_\theta$ of the measure $Q$, which is defined as
\begin{equation}
\label{eq:tildeQ}
\frac{d\bar{Q}}{dP} =\frac{\mbox{e}^{-\theta X}}{E[\mbox{e}^{-\theta X}]} = \mbox{e}^{-\theta X -
\tilde{\psi}(\theta)},
\end{equation}
where $\tilde{\psi}(\theta)$ is $\log\tilde{\Psi}(\theta)$ with $\tilde{\Psi}
(\theta)=E[\mbox{e}^{-\theta X}]$.

To present a connection between $\bar{Q}$ and $Q$, we consider their probability densities with respect
to Lebesgue measure $\mathcal{L}$. It is straightforward to see
that $\tilde{\psi}(\theta)= \psi(-\theta),$ which implies
\begin{equation}\label{eq:tildeQa}
\frac{d\bar{Q}_\theta}{d\mathcal{L}} = \mbox{e}^{-\theta x -\tilde{\psi}(\theta)} \frac{dP}{d\mathcal{L}} = \mbox{e}^{-\theta x -{\psi}(-\theta)} \frac{dP}{d\mathcal{L}} =
\mbox{e}^{ (-\theta) x -{\psi}(-\theta)} \frac{dP}{d\mathcal{L}} = \frac{dQ_{-\theta}}{d\mathcal{L}}.
\end{equation}

The following theorem states the existence and uniqueness for the optimization procedure
(\ref{optimization}), and provides a simplification on the RHS of (\ref{abeq}) using $\bar{Q}$.
Before that, we need a condition for $\Psi(\theta)$ being steep to ensure the finiteness of the moment generating function
$\Psi(\theta)$. To define steepness,  let $\theta_{\max} := \sup \{ \theta: \Psi(\theta) < \infty \}$
(for light-tailed distributions, we have $0 < \theta_{\max} \leq \infty$). Then steepness means $\Psi(\theta) \to \infty$ as $\theta \to \theta_{\max}$.

\begin{theorem}
\label{optimality}
Suppose the moment generating function $\Psi(\theta)$ of $X$
exists and second order continuously differentiable for $\theta \in {\bf I} \subset {\bf R}$. Furthermore, assume that $\Psi(\theta)$ is steep and $E[X|X\in A] > E(X):=\mu$.
Then there exists a unique solution for the optimization problem {\rm (\ref{optimization})}, which
satisfies
\begin{equation}
\label{eq:theta_ast}
\psi'(\theta) = E_{\bar{Q}_\theta}[X|X\in A].
\end{equation}
\end{theorem}

The proof of Theorem \ref{optimality} will be given in the Appendix.

In summary, the three-step procedures to find $\theta^\ast$ are
\begin{enumerate}
\item Calculate the cumulant generating function $\psi(\theta)$ of $X$ and its derivative
$\psi'(\theta)$.
\item Find the exponential tilting measure $Q$.
\item Find $\theta^\ast$ as the solution of $\psi'(\theta)=E_{\bar{Q}_\theta}[X|X\in A]$.
\end{enumerate}

\noindent
{\bf Remark 1:} Note that the optimal tilting point $\theta^\ast$ obtained in (\ref{eq:theta_ast}) highlights the fact that the tilting probability measure
depends on the likelihood ratio (or the embedding probability $Q$ in terms of $\bar{Q}$), the region, and the statistics of interest.
Simple cases such as normal distribution and $t$-distribution
are in Fuh and Hu (2004), and Fuh et al. (2011), respectively, by using the technique of change of variables. The idea of
using conjugate measure in Theorem \ref{optimality} seems to be new and simple according to our best knowledge.

\noindent
{\bf Remark 2:} In addition, characterization (\ref{eq:theta_ast}) is easy to implement as will be illustrated in Sections 3 and 4.
Furthermore, it provides an insightful interpretation of $\theta^\ast$, which
can be used to compare with the large deviation tilting.
To be more specific, when $A=\{w: X(w) > a\}$ for $a > \mu$, it is known that $\theta^\ast$ satisfies $\psi'(\theta)= E_{Q_\theta} X$.
(\ref{eq:theta_ast}) indicates that the optimal tilting $\theta^\ast$ satisfies
\begin{eqnarray}\label{com}
E_{Q_\theta} X = E_{\bar{Q}_\theta}[X|X >a] > a.
\end{eqnarray}
A comparison between (\ref{com}) with (\ref{g4}) shows that the large deviation tilting parameter is the dominating point
$a$; while the optimal tilting $\theta^\ast$ is inside the region of  $\{w: X(w) > a\}$.
Similar interpretation can be applied to the case of $A$ is a convex set in ${\bf R}^q$.
However, when $A=\{w: X(w) \in (-a,a)^c\}$,
where $c$ denotes the complement, it is nature to spit $A=\{w: X(w) > a\} \cup \{w: X(w) < -a\}$
and apply importance sampling for each part. A simple guidance for the stratification, based on relative
efficiency, can be found in Fuh and Hu (2004).


\noindent
{\bf Remark 3:} To approximate $z=EZ$ where $Z \geq 0$ $P$-a.s.. Define $P^* (dw)= \frac{|Z|} {E|Z|}
P(dw)$ and $L^* = \frac{E|Z|}{|Z|}$.
It is known (cf. Theorem 1.2 in Chapter V, Asmussen and Glynn, 2007) that the importance sampling
estimator $ZL^*$ under $P^*$ has zero variance. By using
similar idea, tilting probability obtained by (\ref{eq:theta_ast}) can be regarded as an approximation of
the zero variance importance sampling estimator within a parametric class.

\subsection{Calculating the optimal alternative distribution}

Before employing the importance sampling method, it is first
necessary to identify the optimal alternative distribution. Since
the optimal $\theta$ in (\ref{eq:theta_ast}) cannot be computed directly, to
find $\theta$ in the third step, we consider a simple recursive
procedure for a general equation,
\begin{equation}\label{eqn:g=f}
g(\theta) = h(\theta),
\end{equation}
for some functions $g(\theta)$ and $h(\theta)$. In our setting,
$g(\theta)=\psi'(\theta)$ and $h(\theta)=E_{\bar{Q}}[X|X\in A]$.

A simple recursive procedure is implemented as follows.
\begin{enumerate}
\item Start with an arbitrary $\theta^{0}$. Set $i=1$.
\item Calculate $t^i = g(\theta^{i-1})$ and find $\theta^i$ as the solution of $h(\theta) = t^i$.
\item Set $i=i+1$. Return to 2 until $(\theta^{i+1}-\theta^{i})/\theta^{i}$ becomes very small.
\end{enumerate}

Because the exact objective function $G(\cdot)$ is difficult to
optimize directly, the proposed method
replaces $G$ with a quasi objective function and does optimization on
the latter. In the preceding algorithm, the initial value $\theta^{(0)}$ can be
chosen as a {\it dominating} point of the event $\{X \in A \}$.
Furthermore, if $\theta^{(0)}$ is sufficiently large, the density of
$X$ decreases rapidly and the solution of (\ref{abeq}) is close
to $\theta^{(0)}$. Therefore, fast convergence of the recursive
algorithm is to be expected.



Let $\theta^*$ be the solution to (\ref{eqn:g=f}), i.e., $g(\theta^*)= h(\theta^*)$.
By using an argument similar to Theorem 2 of Fuh et al. (2011), we have the following proposition.
\begin{proposition}\label{recursion}
Choose a dominating point of the event $\{X \in A \}$ as the
initial value $\theta^{(0)}$. Then

{\rm i)} the recursive algorithm either converges to $\theta^*$ or
alternates in the limit between a pair of values
$\underline{\theta}\neq\bar{\theta}$ satisfying
\begin{equation}\label{alternate}
g(\underline{\theta})=h(\bar{\theta})~~{\rm and}~~ h(\underline{\theta})=g(\bar{\theta}).
\end{equation}

{\rm ii)} If there does not exist $\underline{\theta}\neq\bar{\theta}$
such that {\rm (\ref{alternate})} holds, then the recursive
algorithm converges to the solution of {\rm (\ref{eqn:g=f})}.
\end{proposition}
In this subsection, we introduce a simple recursive algorithm. Alternative root-finding algorithms such as bisection method,
Newton's method, secant method, etc. can also be used to find the root in (\ref{eqn:g=f}).

\subsection{Approximating optimal tilting probability measure}

In this subsection, we get the optimal tilting probability measure by
approximating $\psi'(\theta)$ for a moderate deviation event.
Let $X_1,\ldots,X_n$ be $\mbox{i.i.d.}$  random
variables from a distribution function $F$, with mean
$\mu$ and variance $\sigma^2$. We want to estimate by simulation of
the probability
\begin{equation}\label{probab}
P(S_n / n \leq a_n),
\end{equation}
for some $a_n$, where $S_n= \sum_{i=1}^n X_i.$
Note that the probability is small when $a_n - \mu < 0,$ and $a_n - \mu =O(1)$.
Assume that the moment generating function of $X_1$ exists for some $\theta$ belonging
to some interval $\Theta$ which contains the origin, and let $\psi (\theta):=\log E(\exp(\theta X_1))$
denote the cumulant generating function of $X_1$. Note that $\mu = \psi'(0)$,
$\sigma^2 = \psi''(0)$. By using the technique developed in Section 2.1, we first embed
the original probability $P$ in the following exponential family
\begin{eqnarray}\label{embd}
 \frac{d\bar{Q}_\theta} {d\mathcal{L}}(x) = \exp (\theta x - \psi(\theta))\frac{dP}{d\mathcal{L}} (x).
\end{eqnarray}
The class of estimators considered for the probability defined in (\ref{probab}) is
\begin{eqnarray}\label{embdlik}
\delta_n=I(S_n/n \leq a_n) \prod_{j=1}^n
 \frac{dP(X_j)}{d Q_{\theta}(X_j)},
\end{eqnarray}
where $\{X_j\}$ are distributed from $Q_\theta$.
By Theorem 1, we have that if $\mu =0$ and
$a_n =a < 0$, the optimal point $\theta^{\ast}$ in simulations is given by
$\psi'(\theta^{\ast})= E_{\bar{Q}(\theta^\ast)}[X|X > a]$. Or $\psi'(\theta^{\ast})= a$ for
large deviation tilting.

The starting point of the approximation is that $a_n-\mu=o(1)$,
or that the probability (\ref{probab}) is not exceedingly small. For instance one may
be interested in the probability $P(S_n-n \mu \le
a\sigma\sqrt{n})$ for $a < 0$,
and this leads to $a_n = \mu + a\sigma / \sqrt{n}$.
The point $\theta^{\ast}$
can be approximated in the following manner by a one-term Taylor expansion
of $\psi'(\theta^{\ast})$:
\begin{equation}
\psi'(0) + \theta^{\ast} \psi''(0) \cong E_{\bar{Q}_{\theta^\ast}}[X|X > a_n]
\Longrightarrow \mu + \sigma^2 \theta^{\ast}  \cong E_{\bar{Q}_{\theta^\ast}}[X|X > a_n].
\end{equation}
The calculations above suggest that when $a_n-\mu \rightarrow 0$,
expansions for (\ref{embd}) and the optimal choice of parameter $\theta^{\ast}$ can
be obtained through the first few moments of $F$ (more specifically, through
the mean and variance) without appealing to the properties of moment
generating function. Therefore it is illuminating to consider a slightly
different approach to the problem of estimating (\ref{probab}), with $a_n=\mu +
a \sigma/\sqrt{n}$, especially, for heavy-tailed distributions. An alternative approach
for heavy-tailed distribution without moment generating function is via the
method of transform random variables, cf. Asmussen and Glynn (2007).

Example: Pareto distribution. Let $f(x) = \alpha (1 + x)^{-\alpha - 1}$ and $\tilde{f}(x)
 = \tilde{\alpha} (1 + x)^{-\tilde{\alpha} - 1}$, where $\tilde{\alpha} \to 0$.  Here we consider
$ \tilde{\alpha}= \alpha \theta(a)$, and $\theta(a)$ is a solution of
\begin{equation}
\mu + \sigma^2 \theta  = E_{\bar{Q}_\theta}[X|X > a_n].
\end{equation}

\noindent
{\bf Remark 4:}
Note that in (\ref{embd}), we apply the idea of exponential embedding for a non-parametric
distribution $F$.
Further approximation can be done on (\ref{embdlik}) via the LAN family.
This idea had been carried out in the bootstrap setting of Fuh and Hu (2004, 2007).
More detailed analysis along this line and the comparison of non-paremetic importance sampling in
Zhang (1996) and Neddermeyer (2009), for instance, will be published in a separate paper.


\section{Examples and Simulation Study}
\def\theequation{3.\arabic{equation}}
\setcounter{equation}{0}

\subsection{Examples}
To illustrate the general account of importance sampling, in this
subsection, we study three examples: normal distribution, noncentral $\chi^2$ distribution
and compound Poisson processes, and report several other interesting distributions.
The simulation event is $\{X>a\}$ for a given random variable $X$ and some $a > 0$.
In these examples, we explicitly calculate a closed-form formula of $E_{\bar{Q}}[X|X\in A]$ when possible.

\subsection*{Example 1: Normal distribution}

Let $X$ be a random variable with standard normal distribution,
denoted by $N(0,1)$, with probability density function (pdf)
$\frac{dP}{d\mathcal{L}}=e^{-{x^2}/{2}}/\sqrt{2\pi}.$
Standard calculation gives $\Psi(\theta) = \text{e}^{\theta^2/2}$,
$\psi(\theta) = \theta^2/2$ and $\psi'(\theta) = \theta$. In this
case, the tilting measure $Q$ is $N(\theta, 1)$, a location shift,
and $\bar{Q}$ is $N(-\theta,1)$. Applying Theorem \ref{optimality}
and using the fact that $X|\{X>a\}$ is a truncated normal
distribution with minimum value $a$ under $\bar{Q}$, $\theta^\ast$
needs to satisfy
\begin{eqnarray}
\theta =\frac{\phi(a+\theta)}{1-\Phi(a+\theta)}-\theta. \label{normaltilt}
\end{eqnarray}

Alternatively, $G(\theta) = \text{e}^{\theta^2}(1-\Phi(a+\theta))$, and $\theta^\ast$ must
satisfy the first-order condition, $2\theta
(1-\Phi(a+\theta)) = \phi(a+\theta)$, or equivalently,
$\theta = {\phi(a+\theta)}/{2(1-\Phi(a+\theta))}$.
By using $\frac{1-\Phi(x)}{\phi(x)} \sim \frac{1}{x}$ as $x \to \infty$,
it is easy to see from equation (\ref{normaltilt}) that $\theta^\ast \sim a$ when $a$ is large.
This is the same as the large deviation tilting probability.

We remark the normal distribution has been analysed in Fuh and
Hu (2004), for illustration, we consider this example from our
general account and provide a simple and explicit characterization for $\theta^\ast$,
in the sense that the right-hand-side of (\ref{normaltilt}) is a straightforward application of Theorem \ref{optimality}.

\subsection*{Example 2: Noncentral $\chi^2(\lambda,\kappa)$ distribution}

Let $Z_i$ be independent, normally distributed random variables
with mean $\mu_i$ and variances $\sigma_i^2$, for $i=1,\ldots, \kappa$.
Then the random variable
$X = \sum_{i=1}^\kappa \left(\frac{Z_i}{\sigma_i}\right)^2$
is distributed according to the noncentral $\chi^2$ distribution. It
has two parameters: $\kappa$ which specifies the number of degrees of
freedom, and $\lambda$, the noncentrality parameter, is defined as
$\lambda =\sum_{i=1}^\kappa \left(\frac{\mu_i}{\sigma_i}\right)^2.$

The pdf of $X$ is
\begin{eqnarray}\label{nchi2a}
\frac{dP}{d\mathcal{L}} = \frac{1}{2} e^{-(x+\lambda)/2} \bigg(\frac{x}{\lambda}\bigg)^{\kappa/4 - 1/2}
I_{\kappa/2 -1}(\sqrt{\lambda x}),\quad x\geq 0,
\end{eqnarray}
where $I_v(z)$ is a modified Bessel function of the first kind given by
\[ I_v(z)= (z/2)^v \sum_{j=0}^\infty \frac{(z^2/4)^j}{j! \Gamma(v+j+1)}. \]
Alternatively, the pdf of $X$ can be written as
\begin{eqnarray}\label{nchi2}
\frac{dP}{d\mathcal{L}} =\sum_{i=0}^\infty \frac{e^{-\lambda/2} (\lambda/2)^i}{i!} f_{Y_{\kappa+2i}}(x),\quad x\geq 0,
\end{eqnarray}
where $Y_q$ is a chi-square distribution with $q$ degrees of freedom, denoted by $Y_q\sim \chi^2(q)$.
By representation (\ref{nchi2}),
the noncentral $\chi^2$ distribution is a Poisson
weighted mixture of central $\chi^2$ distributions. Suppose that a
random variable $N$ has a Poisson distribution with mean
$\lambda/2$, and the conditional distribution of $X$ given $N=i$ is
$\chi^2$ with $\kappa+2i$ degrees of freedom. Then the unconditional
distribution of $X$ is noncentral $\chi^2$ with $\kappa$ degrees of
freedom, and noncentrality parameter $\lambda$, or,
$$X|\{N=i\}\sim \chi^2(\kappa+2i),\quad N\sim Pois(\lambda/2).$$

Let the likelihood ratio
$\frac{dQ}{dP}=e^{\theta X-\psi(\theta)},$
where $\psi(\theta)= \log E[\exp(\theta X)]$. To derive the tilting formula of $X$, note that
$\Psi(\theta) = {\exp\left(\frac{\lambda\theta}{1-2\theta}\right)}/{(1-2\theta)^{\kappa/2}},$
$\psi(\theta) = \frac{\lambda\theta}{1-2\theta} -\frac{\kappa}{2}\log(1-2\theta),$
and $\psi'(\theta) = \frac{\lambda + \kappa (1-2\theta)}{(1-2\theta)^2}.$
Therefore, the exponential tilting measure $Q$ becomes
\begin{eqnarray}\label{ncctilting}
\displaystyle \frac{dQ}{d\mathcal{L}}&=& \frac{dQ}{dP}\frac{dP}{d\mathcal{L}}
=e^{\theta x-\psi(\theta)}\sum_{i=0}^\infty \frac{e^{-\frac{\lambda}{2}}(\frac{\lambda}{2})^i}{i!} \frac{1}
{\Gamma(\frac{\kappa+2 i}{2})2^{\frac{\kappa+2 i}{2}}}x^{\frac{\kappa+2 i}{2}-1}e^{-\frac{x}{2}} \\
&=&\sum_{i=0}^\infty \frac{e^{-\frac{\lambda}{2(1-2\theta)}}(\frac{\lambda}{2(1-2\theta)})^i}{i!} \frac{1}
{\Gamma(\frac{\kappa+2 i}{2})(\frac{2}{1-2\theta})^{\frac{\kappa+2 i}{2}}}x^{\frac{\kappa+2 i}
{2}-1}e^{-\frac{x}{\frac{2}{1-2\theta}}}. \nonumber
\end{eqnarray}
Hence, by (\ref{ncctilting}) $X$ can be characterized by
$X|\{N=i\}\sim \Gamma\left(\frac{\kappa+2 i}{2},\frac{2}{1-2\theta}\right)$ with $N \sim Pois\left(\frac{\lambda}{2(1-2\theta)}\right)$
under $Q$. By Theorem \ref{optimality}, $\theta^\ast$ is the solution of the following equation
\begin{eqnarray}
\frac{\lambda + \kappa(1-2\theta)}{(1-2\theta)^2}=E_{\bar{Q}}[X |X>a], \label{nonchi2}
\end{eqnarray}
where $X$ follows
$X|\{N=i\}\sim \Gamma\left(\frac{\kappa+2 i}{2},\frac{2}{1+2\theta}\right)$ with  $N\sim Pois\left(\frac{\lambda}{2(1+2\theta)}\right)$
under $\bar{Q}$.

When $\lambda = 0$, this reduces to $\chi^2(\kappa)$ distribution and the optimal $\theta^\ast$ needs to satisfy
\begin{eqnarray}
\frac{\kappa}{1-2\theta}=E_{\bar{Q}}[X|X > a]. \label{cenchi2}
\end{eqnarray}When the degree of freedom $\kappa$ equals 2, $\chi^2(2)$ distribution reduces to the exponential distribution with mean 2. By the memoryless property, we have an explicit formula for $C(\theta)$ and hence $\theta^\ast$ needs to satisfy
$\frac{2}{1-2\theta} = a + \frac{2}{1+\theta}.$
As a result, we have an explicit solution for
\begin{eqnarray}\label{expon}
\theta^\ast = \frac{-1+\sqrt{1+a^2}}{a} < 1.
\end{eqnarray}
Note that the tilting formula (\ref{expon}) can also be found in Example 1 on p.22 of
L'Ecuyer, Mandjes and Tuffin (2009).

\subsection*{Example 3: Compound Poisson process}

Let $R_t=\sum_{n=1}^{N(t)} \log Y_n$ be the compound Poisson process,
where jump event $N(t)$ is assumed to
follow a Poisson process with parameter $\lambda$, and the
jump sizes $Y_n$ are assumed to follow a lognormal distribution
with parameters of location $\eta$ and scale $\delta^2$.
Let $A=\{R_t>r_p\}$ for a given $r_p$, and compute
\begin{eqnarray*}
P(R_t>r_p)
&=&\sum_{n=0}^\infty \frac{e^{-\lambda t} (\lambda t)^n }{n!}P(\sum_{i=1}^n Z_i>r_p|N(t)=n)\\
&=&\sum_{n=0}^\infty \frac{e^{-\lambda t} (\lambda t)^n }{n!}P(Z>\frac{r_p-n\eta}{\sqrt{n \delta^2}}|N(t)=n),\\
\end{eqnarray*}
where $Z_i$, $i=1,\ldots,n$, are i.i.d. normal with mean $\eta$ and variance $\delta^2$ and $Z=\frac{\sum_{i=1}^n Z_i-n\eta}{\sqrt{n \delta^2}}$.
Denote $f(R_t)= R_t- r_p$, and let the likelihood ratio
\begin{eqnarray}\label{cpm}
\frac{dQ}{dP}=e^{\theta f(R_t)-\psi(\theta)},
\end{eqnarray}
where $\psi(\theta)=\log E[\exp(\theta f(R_t))]$.

Here the original measure $P$ under $N(t)\sim Pois(\lambda t)$ and $\log Y_n \sim
N(\eta,\delta^2)$ becomes
\[
\frac{dP}{d\mathcal{L}}= \frac{e^{-\lambda t} (\lambda t)^n }{n!}(\frac{1}{\sqrt{2\pi \delta^2}})^n 
e^{-\frac{\sum_{i=1}^n (z_i-\eta)^2}{2 \delta^2}}.
\]
and
\begin{eqnarray*}
\displaystyle E[e^{\theta f(R_t)}]
&=&e^{-\theta r_p}E[e^{\theta \sum_{i=1}^{N(t)}Z_i}]
=e^{-\theta r_p}\sum_{n=1}^\infty \frac{e^{-\lambda t} (\lambda t)^n }{n!}E[e^{\theta \sum_{i=1}^n Z_i}|N(t)=n]\\
&=&e^{-\theta r_p}\sum_{n=1}^\infty \frac{e^{-\lambda t} (\lambda t)^n }{n!}e^{(\theta \eta+\frac{1}{2}\theta^2 \delta^2)n}
=e^{\lambda t(e^{\theta \eta+\frac{1}{2}\theta^2 \delta^2}-1)-\theta r_p}.
\end{eqnarray*}
Therefore,
$\psi(\theta)=\lambda t(e^{\theta \eta+\frac{1}{2}\theta^2 \delta^2}-1)-\theta r_p,$
and
$\psi'(\theta)=\lambda te^{\theta \eta+\frac{1}{2}\theta^2 \delta^2}(\eta+\theta \delta^2)-r_p.$
Hence, the likelihood ratio (\ref{cpm}) becomes
\[
\frac{dQ}{dP}=e^{\theta \sum_{i=1}^{n}z_i}e^{-\lambda t(e^{\theta \eta+\frac{1}{2}\theta^2 \delta^2}-1)}.
\]For a given $N(t)=n$,
\begin{eqnarray*}
\frac{dQ}{d\mathcal{L}}&=& \frac{dQ}{dP} \frac{dP}{d\mathcal{L}}
= e^{\theta \sum_{i=1}^{n}z_i}e^{-\lambda t(e^{\theta \eta+\frac{1}{2}\theta^2 \delta^2}-1)}
\frac{e^{-\lambda t} (\lambda t)^n }{n!}(\frac{1}{\sqrt{2\pi \delta^2}})^n e^{-\frac{\sum_{i=1}^n (z_i-\eta)^2}{2 \delta^2}} \\
&=&\frac{(\lambda t e^{\theta \eta+\frac{1}{2}\theta^2 \delta^2})^n e^{-\lambda te^{\theta \eta+\frac{1}{2}\theta^2 \delta^2}}}{n!}
(\frac{1}{\sqrt{2\pi \delta^2}})^n e^{-\frac{\sum_{i=1}^n (z_i-(\eta+\theta \delta^2))^2}{2 \delta^2}}.
\end{eqnarray*}As a result, $R_t=\sum_{i=1}^{N(t)}Z_i$ with $Z_i \sim N(\eta+\theta \delta^2, \delta^2)$ and $N(t)\sim Pois(\lambda e^{\theta \eta + \theta^2\delta^2/2}t)$
under $Q$. By Theorem \ref{optimality}, $\theta^\ast$ needs to satisfy
$$
\lambda te^{\theta \eta+\frac{1}{2}\theta^2 \delta^2}(\eta+\theta \delta^2)-r_p = E_{\bar{Q}}\left[R_t|R_t>r_p\right],
$$where $R_t=\sum_{i=1}^{N(t)}Z_i$ with $R_t|\{N(t)=n\}\sim N(\eta-\theta \delta^2, \delta^2)$ and $N(t)\sim Pois(\lambda e^{ -\theta \eta + \theta^2\delta^2/2}t )$
under $\bar{Q}$.

\begin{table}\label{T1}

Table 1 reports importance sampling tilting probability for eight distributions:
Binomial, Poisson, exponential, normal, $\chi^2$, Gamman, noncentral $\chi^2$, and uniform. The
case of $t$ distribution can be found in Fuh et al. (2011). This table includes the family of
tilting probability $Q$ and $\psi'(\theta)$.
Note that in Table 1, when $A=\{X> a\}$ for some $a > 0$, then $E_{\bar{Q}}[X|X >a]=a+\frac{1}{1+\theta}$
for exponential distribution $E(1)$ and  $=\frac{\phi(a+\theta)}{1-\Phi(a+\theta)}-\theta$ for
standard normal distribution.
By Theorem \ref{optimality}, we have that $\bar{Q}=Q_{-\theta}$ in Examples 1--7.

\caption{Summary of distributions and their tilting measures.}
\begin{center}
\begin{tabular}{lllc}\toprule
Ex & $P$ & $Q$ &  $\psi'(\theta)$  \\ \midrule\vspace{1cm}
1 & $B(n,p)$ & $B(n, \frac{p\text{e}^\theta}{p\text{e}^\theta+(1-p)})$ & $\frac{np\text{e}^{\theta}}{1-
p+p\text{e}^\theta}$ \\\vspace{1cm}
2 & $Pois(\lambda) $ & $Pois(\lambda \text{e}^{\theta})$ &  $\lambda\text{e}^\theta$ \\\vspace{1cm}
3 & $N(0,\sigma^2)$ & $N(\theta,\sigma^2)$ &  $\theta \sigma^2$  \\\vspace{1cm}
4 & $E(1)$ & $E(\frac{1}{1-\theta})$ &  $\frac{1}{1-\theta}$\\\vspace{1cm}
5 & $\chi^2(\kappa)$ & $\Gamma(\frac{\kappa}{2}, \frac{2}{1-2\theta})$ & $\frac{\kappa}{1-2\theta}$\\\vspace{1cm}
6 & $\Gamma(\alpha, \beta)$ & $\Gamma(\alpha,\frac{\beta}{1-\beta\theta})$ &  $\frac{\alpha\beta}{1-\beta\theta}$ \\\vspace{1cm}
7 & $NC\chi^2(\kappa, \lambda)$ & $X|\{N = i\} \sim \Gamma(\frac{\kappa +2i}{2},\frac{2}{1-2\theta})$, $N\sim Pois(\frac{\lambda}{2(1-2\theta)})$   & $\frac{\lambda+ \kappa(1-2\theta)}{(1-2\theta)^2}$\\\vspace{1cm}
8 & $Uniform(0,1)$ & $\propto\text{e}^{\theta X}\textbf{1}_{\{0\leq X\leq 1\}}$ &$\frac{\text{e}^{\theta}}{\text{e}^\theta
-1}-\frac{1}{\theta}$ \\\bottomrule
\end{tabular}
\end{center}
\end{table}

\pagebreak

\subsection{Simulation Study}

In this subsection, we present some numerical results on relative efficiency using the
method outlined in Theorem 1 for estimating tail probabilities of  $p = P\{X >a\}$ for some
$a > 0$. The relative efficiency is defined as the
variance ratio between the crude Monte Carlo estimator and the importance sampling estimator.

Table 2 compares the relative efficiency (RE) for various distributions. Here $a$ is chosen such
that $p=0.1$ (central event), $0.05,~0.01$ ($95\%,~99\%$ confidence interval),
$0.001$ (VaR type probability) and $0.0001$ (rare event). Monte Carlo sample size is $10,000$ for
each case. Note that the relative efficiency in normal case is the largest.

Simulations are conducted for central $\chi^2(\kappa),$ noncentral $\chi^2(\lambda,\kappa)$
and gamma $\Gamma(\alpha, \beta)$ distributions.
Since the results are almost the same, we only report the simulation results
for noncentral $\chi^2(\lambda,\kappa)$ distribution
in Table 3.  We can see that the proposed method is significantly more efficient than the
naive Monte Carlo in $10,000$ simulations for all probabilities of
$p=0.1,~0.05,~0.01,~0.001$ and $0.0001$.
Furthermore, the efficiency gain is larger for smaller probabilities against the naive method.
The normal case can be found in Fuh and Hu (2004), which will be used, along with its square and
compound Poisson model, for VaR computation.
Normal distribution and $\chi^2$ distribution will be used for
importance resampling in bootstrapping confidence region of parameters in regression model.

\begin{table}
\caption{Relative efficiency for different distributions. }
\begin{center}
\begin{tabular}{lrrrrr}\toprule
$p$  &  $N(0,1)$     &  $E(1)$   &  $\chi^2(1)$  &  $\Gamma(4,10)$   &  $\chi^2(2, 10)$  \\\midrule
0.0001   &  2409.74      &  818.53   &  603.61   &  1282.87      &  1529.46      \\
0.001    &  290.90   &  109.88   &  82.74    &  166.00   &  192.20   \\
0.01     &  38.06    &  16.57    &  12.90    &  23.74    &  26.54    \\
0.05     &  9.98     &  4.99     &  4.04     &  6.76     &  7.34     \\
0.1  &  5.77     &  3.13     &  2.60     &  4.10     &  4.38     \\
\bottomrule
\end{tabular}
\end{center}
\end{table}


\begin{table}
\caption{We compare the performance, in terms of means and standard errors (SE), of estimating the probability $P(NC\chi^2(\kappa, \lambda)>a)$ using the naive Monte Carlo method (MC) and our importance sampling method (IS) for a combination of parameters ($\kappa$, $\lambda$) and a variety of boundary values $a$ with a sample size of 10,000. $p$ is the theoretical value of the probability of interest. $\theta^*$ is the optimal tilting parameter. Both the relative efficiency (RE) and the benchmark RE (RE*) are reported, where the former is calculated as the ratio of the variance of the Monte Carlo estimate over that of the importance sampling estimate, and the later is calculated as $p(1-p)/(G(\theta^*)-p^2)$. }
\begin{center}
\begin{tabular}{llrrrrrrrr}\toprule
($\kappa$, $\lambda$) & $p$ & $a$ & $\theta^*$ & \multicolumn{2}{c}{MC} & \multicolumn{2}{c}{IS} & RE & RE* \\
\cmidrule(l){5-6} \cmidrule(l){7-8}
& & & & mean & SE & mean & SE & & \\\midrule
(2, 1)& 0.0001 & 24.28 & 0.38 & 0.000000 & 0.000000 & 0.000097 & 0.000003 &   0.00 & 1059.49 \\
& 0.001 & 18.65 & 0.36 & 0.001200 & 0.000346 & 0.000981 & 0.000027 & 168.64 & 136.82 \\
& 0.01 & 12.85 & 0.33 & 0.010300 & 0.001010 & 0.009940 & 0.000226 &  20.04 &  19.51 \\
& 0.05 & 8.64 & 0.29 & 0.050100 & 0.002182 & 0.049579 & 0.000913 &   5.71 &   5.59 \\
& 0.1 & 6.77 & 0.26 & 0.101400 & 0.003019 & 0.099657 & 0.001610 &   3.52 &   3.42 \\
\\
(2, 5)& 0.0001 & 37.05 & 0.31 & 0.000200 & 0.000141 & 0.000104 & 0.000003 & 2653.12 & 1381.16 \\
& 0.001 & 29.88 & 0.28 & 0.001200 & 0.000346 & 0.001045 & 0.000025 & 198.95 & 173.94 \\
& 0.01 & 22.23 & 0.25 & 0.009600 & 0.000975 & 0.009987 & 0.000202 &  23.28 &  24.22 \\
& 0.05 & 16.38 & 0.21 & 0.047500 & 0.002127 & 0.049684 & 0.000830 &   6.56 &   6.76 \\
& 0.1 & 13.64 & 0.19 & 0.098600 & 0.002981 & 0.099130 & 0.001493 &   3.99 &   4.07 \\
\\
(5, 1)& 0.0001 & 30.02 & 0.36 & 0.000000 & 0.000000 & 0.000101 & 0.000003 &   0.00 & 1213.85 \\
& 0.001 & 24.07 & 0.34 & 0.001000 & 0.000316 & 0.001030 & 0.000026 & 152.60 & 156.69 \\
& 0.01 & 17.83 & 0.30 & 0.009800 & 0.000985 & 0.009918 & 0.000209 &  22.13 &  22.30 \\
& 0.05 & 13.17 & 0.26 & 0.049000 & 0.002159 & 0.049644 & 0.000856 &   6.36 &   6.37 \\
& 0.1 & 11.03 & 0.23 & 0.101400 & 0.003019 & 0.099369 & 0.001522 &   3.93 &   3.89 \\
\\
(5, 5)& 0.0001 & 41.73 & 0.30 & 0.000100 & 0.000100 & 0.000101 & 0.000003 & 1401.51 & 1434.04 \\
& 0.001 & 34.35 & 0.28 & 0.000600 & 0.000245 & 0.001008 & 0.000024 & 106.09 & 181.10 \\
& 0.01 & 26.42 & 0.24 & 0.011800 & 0.001080 & 0.010086 & 0.000200 &  29.05 &  25.28 \\
& 0.05 & 20.29 & 0.20 & 0.046300 & 0.002101 & 0.048860 & 0.000805 &   6.81 &   7.05 \\
& 0.1 & 17.38 & 0.18 & 0.093900 & 0.002917 & 0.098820 & 0.001448 &   4.06 &   4.22 \\
\bottomrule
\end{tabular}
\end{center}
\end{table}

\clearpage


\section{Applications}
\def\theequation{4.\arabic{equation}}
\setcounter{equation}{0}

To illustrate the applicability of our proposed tilting formula, we present
two applications in this section. In subsection 4.1, we apply the importance sampling
for portfolio VaR computation, to which the tilting formula is given under the multivariate
jump diffusion model of the underlying assets. In subsection 4.2, we study importance
resampling  for bootstrapping confidence regions in regression models.
By using the idea of pseudo-maximum likelihood estimator (PMLE) and the
criterion of minimizing variance of the Monte Carlo estimator
under a parametric family, we propose a titling formula based on the
parametric family of normal (or $\chi^2$) distributions.

\subsection{Evaluating Value at Risk}

As a standard benchmark for market risk disclosure, VaR is the loss
in market value over a specified time horizon that will not be
exceeded with probability $1-p$. Hence define VaR as the quantile
$l_p$ of the loss $L$ in portfolio value during a holding period of
a given time horizon $\Delta t$. To be more specific, we express the
portfolio value $V(t,S(t))$ as a function of risk factors and time,
where $S(t)=(S^{(1)}(t),\ldots,S^{(d)}(t))^{T}$ comprises the $d$
risk factors to which the portfolio is exposed at time $t$ and
$T$ denotes the transpose of a matrix. The loss of the portfolio
over the time interval $[t,t+\Delta{t}]$ is
$L= V(t, S(t))-V(t+\Delta t,S(t+\Delta t)).$
Therefore VaR, $l_{p}$, associated with a given probability $p$ and
time horizon $\Delta t$, is given by
\begin{eqnarray}
{P}(L>l_{p})=p .  \label{def of VaR}
\end{eqnarray}

Assume $S(t)$ follows a $d$-dimensional jump diffusion model such that
the return processes are described by the stochastic differential equations
\begin{eqnarray}\label{jdm}
r_t^{(i)}=\displaystyle\frac{dS^{(i)}(t)}{S^{(i)}(t)}=\mu^{(i)} dt+ \sigma^{(i)} dW^{(i)}(t)+\displaystyle\sum_{j=1}^{N(t)}\log Y_j^{(i)},~ i=1,2,\ldots,d,
\end{eqnarray}
where ($W^{(1)}(t)$,\ldots,$W^{(d)}(t)$) is a standard Brownian motion in
$R^d$, $N(t) \sim Pois(\lambda dt)$ is a Poisson process, $\log Y_j^{(i)}$
are independent and identically distributed (i.i.d.) random variables with
$N(\eta^{(i)},{\delta^{(i)}}^2)$ distribution. Here
the drift parameters $\mu^{(i)}$, volatility parameters
$\sigma^{(i)}$, jump frequency $\lambda$, and jump size parameters
$\eta^{(i)}$, $\delta^{(i)}$ are given. Furthermore, we assume that
the Brownian motion and the Poisson process are independent. Denote $\log Y_j \sim
N(\displaystyle\underline{\eta},\underline{\delta}^T
\Sigma_J \displaystyle\underline{\delta})$, where
\[
\displaystyle\underline{\eta}=
\left[
\begin{array}{c}
\eta^{(1)} \\
\eta^{(2)} \\
\vdots \\
\eta^{(d)}
\end{array}
\right],~~~
\displaystyle\underline{\delta}=
\left[
\begin{array}{cccc}
\delta^{(1)} & 0 &\cdots & 0 \\
0& \delta^{(2)}& \cdots & 0 \\
\vdots & \vdots & \ddots & \vdots \\
0 & \cdots &\cdots &\delta^{(d)}
\end{array}
\right],
\Sigma_J= \left[
\begin{array}{cccc}
1& \rho_{12}^J &\ldots & \rho_{1d}^J\\
\rho_{12}^J & 1 & \ldots & \rho_{2d}^J \\
\vdots & \vdots & \ddots & \vdots \\
\rho_{1d}^J & \ldots & \ldots & 1
\end{array}
\right].
\]

For simple notations, denote
\[
\displaystyle\underline{\mu}=
\left[
\begin{array}{c}
\mu^{(1)}\Delta t \\
\vdots \\
\mu^{(d)}\Delta t
\end{array}
\right],~~~
\displaystyle\underline{\sigma}=
\left[
\begin{array}{c}
\sigma^{(1)}\sqrt{\Delta t} \\
\vdots \\
\sigma^{(d)}\sqrt{\Delta t}
\end{array}
\right],~~~
X=
\left[
\begin{array}{c}
X^{(1)} \\
\vdots \\
X^{(d)}
\end{array}
\right]
\sim MN(
\left[
\begin{array}{c}
0 \\
\vdots \\
0
\end{array}
\right],
\Sigma)
,\]
\[
 \Sigma=
\left[
\begin{array}{cccc}
1& \rho_{12} &\ldots & \rho_{1d}\\
\rho_{12} & 1 & \ldots & \rho_{2d} \\
\vdots & \vdots & \ddots & \vdots \\
\rho_{1d} & \ldots & \ldots & 1
\end{array}
\right],~~~
J=
\left[
\begin{array}{c}
\sum_{j=1}^{N(\Delta t)}\log Y_j^{(1)} \\
\sum_{j=1}^{N(\Delta t)}\log Y_j^{(2)} \\
\vdots \\
\sum_{j=1}^{N(\Delta t)}\log Y_j^{(d)}
\end{array}
\right].
\]

A discrete version of (\ref{jdm}) is
\begin{eqnarray}\label{djdm}
r_t^{(i)}=\mu^{(i)} \Delta t+ \sigma^{(i)} \sqrt{\Delta t} X^{(i)}+
\displaystyle\sum_{j=1}^{N(\Delta t)}\log Y_j^{(i)},~~i=1,\ldots,d.
\end{eqnarray}

Next we shall describe a quadratic approximation to the loss $L$.
Let $\underline{r}=(r^{(1)},\ldots,r^{(d)})^T$ be a vector of return of
$d$ assets. Denote $\Delta S = [S(t+\Delta t)-S(t)]\approx \underline{r}$ as the change in $S$ over the
corresponding time interval. The delta-gamma methods
refine the relationship between risk factors and portfolio value by including
quadratic as well as linear terms. The delta-gamma approximation to
the change in portfolio value is
\[V(t+\Delta t,S+\Delta S)-V(t,S)\approx \frac{\partial V}{\partial t}\Delta t+\delta^T \Delta S
+\frac{1}{2}\Delta S^T \Gamma \Delta S,\]
where $\delta=(\delta_1,\ldots,\delta_d)^T$ and $\Gamma=[\Gamma_{ij}]_{i,j=1,\ldots,d}$, with
\[
\delta_i=\frac{\partial V}{\partial S_i},~~~~ \Gamma_{ij}=
\frac{\partial^2 V}{\partial S_i\partial S_j},~~~~i,j=1,\ldots,d,
\]
and all derivatives are evaluated at the initial point $(t,S)$.
Hence we can approximate the loss
\begin{eqnarray}\label{qapprox}
L \approx a_0+a_1^T\underline{r} +\underline{r}^T A_1 \underline{r},
\end{eqnarray}
where $a_0=-\frac{\partial V}{\partial t}\Delta t$ is a scalar,
$a_1=-\delta$ is an $d$-vector and $A_1 = -\Gamma/2$ is a symmetric matrix.

Under the jump diffusion process assumption of $\underline{r}$, we have
\begin{eqnarray}
L  &= &a_0+a_1^T\displaystyle\underline{\mu} +a_1^T \displaystyle\underline{\sigma} X+a_1^T J+
 \displaystyle\underline{\mu}^T A_1 \displaystyle\underline{\mu}+
 \displaystyle\underline{\sigma}^T Z^T A_1 Z\displaystyle\underline{\sigma} \nonumber \\
 &=& b_0+a_1^T \displaystyle\underline{\sigma} X+a_1^T J+
 \displaystyle\underline{\sigma}^T Z^T A_1 X\displaystyle\underline{\sigma},
\end{eqnarray}
where $b_0=a_0+a_1^T\displaystyle\underline{\mu} + \displaystyle\underline{\mu}^T A_1
\displaystyle\underline{\mu}$ and
$J=(\sum_{j=1}^{N(\Delta t)}\log Y_j^{(1)},\ldots,\sum_{j=1}^{N(\Delta t)}\log Y_j^{(d)})^T$.

To have a simple approximation, we neglect the quadratic approximation of the
jump part in (\ref{qapprox}) as it is very small compared to
the return of portfolio. Let $C$ be the square root of the positive definite matrix $\Sigma$
such that $C^TC=\Sigma$. We can
transform the distribution of $X$ into $Z$, which is multivariate normal
distributed with identity covariance matrix,
so that $X= CZ.$
Moreover, $C$ can be chosen so that $C^T A_1 C$ is diagonalized with diagonal elements
$\lambda_1, \ldots, \lambda_d$.
Denote $L_b := L-b_0$, then
\begin{eqnarray}\label{tdiagonal}
L_b &=& a_1^T \displaystyle\underline{\sigma} X+a_1^T J+
 \displaystyle\underline{\sigma}^T X^T A_1 X\displaystyle\underline{\sigma}
 = a_1^T \displaystyle\underline{\sigma} CZ+\displaystyle\underline{\sigma}^T
 Z^TC^TA_1CZ\displaystyle\underline{\sigma} \nonumber+a_1^T J \\
 &=& b^TZ+\displaystyle\underline{\sigma}^T Z^T\Lambda Z\displaystyle\underline{\sigma} +a_1^T
 J=\sum_{j=1}^d b_jZ_j+\lambda_j(\sigma^{(j)})^2\Delta t Z_j^2+a_1^T J \label{Q function},
\end{eqnarray}
where $b^T=a_1^T\displaystyle\underline{\sigma} C$.

By using the tilting formula developed in Theorem \ref{optimality}
of Section 2, we consider a family of
alternative distributions. Let the likelihood ratio of the
alternative probability measures with respect to the target measure
$P$ be of the form
\begin{equation}\label{lr}
\frac{d Q}{d P}=e^{\theta (L_b - r_p) - \psi(\theta)},
\end{equation}
where $L_b$ is defined in (\ref{tdiagonal}), $r_p$ is the $p$th quantile,
and $\psi(\theta)=\log E[\exp(\theta (L_b-r_p))]$ is the cumulant
generating function of $L_b-r_p$ under the target probability measure $P$.
The domain of $\theta$ will be specified after Equation (\ref{lambdaeta}). The
problem of finding the optimal alternative measure is then reduced
to that of identifying the $\theta$-value that yields the minimal
variance for the importance sampling estimator.

We now proceed to find the probability density under the alternative
measure $Q$.  A simple calculation leads that
\begin{eqnarray}\label{psix}
 \psi(\theta) &=& \log E(e^{\theta L_b-\theta r_p}) \\
&=& \frac{1}{2}\sum_{i=1}^d \bigg(\frac{(\theta b_i)^2}{1-2 \theta \lambda_i
(\sigma^{(i)})^2\Delta t}-\log(1-2\theta \lambda_i (\sigma^{(i)})^2\Delta t) \bigg) \nonumber \\
&~& +\lambda \Delta t \bigg(e^{\displaystyle\theta \displaystyle a_1^T\underline{\eta}+\frac{1}{2}
\theta^2 a_1^T\displaystyle(\underline{\delta}^T \Sigma_J
\displaystyle\underline{\delta})a_1}-1\bigg) -\theta r_p, \nonumber
\end{eqnarray}and
\begin{eqnarray}
\nonumber \psi'(\theta)&=&\displaystyle \sum_{i=1}^d
 (\frac{\theta (b_i)^2(1-\theta \lambda_i
(\sigma^{(i)})^2\Delta t)}{(1-2 \theta \lambda_i (\sigma^{(i)})^2\Delta t)^2}+\frac{\lambda_i
(\sigma^{(i)})^2\Delta t}{1-2\theta
\lambda_i(\sigma^{(i)})^2\Delta t})-r_p \\
\label{eq:bootstrap_psiplum}&&+\lambda \Delta t e^{\displaystyle \theta a_1^T \displaystyle\underline{\eta}+\frac{1}{2}
\theta^2 a_1^T \displaystyle(\underline{\delta}^T \Sigma_J
\displaystyle\underline{\delta})a_1}(a_1^T \displaystyle\underline{\eta}+\theta a_1^T
\displaystyle(\underline{\delta}^T \Sigma_J
\displaystyle\underline{\delta}) a_1).
\end{eqnarray}

From (\ref{lr}), and (\ref{psix}), it follows that the importance
sampling is done with an exponential tilting measure.
Therefore, given $N(\Delta t)=n$,
\begin{eqnarray}\label{qdensity}
\displaystyle \frac{d Q}{d\mathcal{L}}&=& \frac{d Q}{dP} \frac{dP}{d\mathcal{L}}
 = \prod_{j=1}^d \displaystyle \frac{1}{\sqrt{2\pi} \sigma_j(\theta)}
\exp\bigg\{-\frac{(z_j-\mu_j(\theta))^2}{2\sigma_j^2(\theta)}\bigg\}
\times \frac{\bigg(\lambda(\theta) \bigg)^n  e^{-\lambda(\theta)}}{n!} \\
&~& \times \bigg(\frac{1}{(2\pi)^{d/2} |\underline{\delta}^T \Sigma_J \underline{\delta}|^{1/2}}\bigg)^n
\exp \bigg\{-\frac{1}{2}\sum_{i=1}^n (v(i)-\eta(\theta))^T (\underline{\delta}^T
\Sigma_J\underline{\delta})^{-1}(v(i)-\eta(\theta)) \bigg\}, \nonumber
\end{eqnarray}
where 
\begin{equation}\label{musigma}
\mu_j(\theta)=\frac{\theta b_j}{1-2 \theta \lambda_j (\sigma^{(i)})^2\Delta t},~~~
\sigma_j^2(\theta)=\frac{1}{1-2 \theta \lambda_j (\sigma^{(i)})^2\Delta t},
\end{equation}
and
\begin{equation}\label{lambdaeta}
\lambda(\theta)=\lambda \Delta t e^{\displaystyle
\theta a_1^T\displaystyle\underline{\eta} + \frac{1}
{2}\theta^2 a_1^T(\displaystyle\underline{\delta}^T \Sigma_J\displaystyle\underline{\delta})a_1},~
\eta(\theta)=\underline{\eta}+\theta a_1^T(\underline{\delta}^T\Sigma_J\underline{\delta}).
\end{equation}
That is, in (\ref{qdensity}) $Z_j, ~j=1, \ldots, d,$ are independent
$N(\mu_j(\theta),\sigma_j^2(\theta))$, and $J_i$ are compound Poisson processes with jump
frequency $\lambda(\theta)$ and jump size $V(i),~i=1,\ldots,n$ has mean $\eta(\theta)$ and variance
matrix $\underline{\delta}^T \Sigma_J\displaystyle\underline{\delta}$.
To guarantee the rational of $Q$ under exponential twisting of measures,
the constant $\theta$ must satisfy $1-2 \theta \lambda_{(1)} (\sigma^{(i)})^2\Delta t>0$ and
$1-2\theta\lambda_{(d)} (\sigma^{(i)})^2\Delta t>0$,
where $\lambda_{(1)}=\max_{1 \leq i \leq d} \lambda_i$ and $\lambda_{(d)}=\min_{1 \leq i \leq d}
\lambda_i$.

To have an efficient importance sampling for approximating $P\{L_b > r_p\}$ for some $r_p > 0$,
we need to characterize the optimal titling $\theta$ via Theorem \ref{optimality}.
Before stating the result, we define some quantities that facilitate the presentation of it.
In view of (\ref{musigma}), define
\begin{equation}\label{alphabar}
\bar{\mu}_j(\theta)=\mu_j(-\theta),~~\bar{\sigma}_j^2(\theta)=\sigma_j^2(-\theta), ~~
\bar{\lambda}(\theta)=\lambda(-\theta),~~ \bar{\eta}(\theta)=\eta(-\theta).
\end{equation}
Let $\bar{V}(j)$ has a $d$-variate normal distribution with mean $\bar{\eta}(\theta)$ and
variance matrix $\underline{\delta}^T\Sigma_J\underline{\delta}$ and $N(\Delta t)$ follows a $Pois(\bar{\lambda}(\theta))$.
Applying Theorem \ref{optimality}, it is straight forward to obtain the following corollary.

\noindent
\begin{corollary}\label{optimaltheta}
Let $\theta$ be such that $1 \pm 2 \theta \lambda_{(1)} (\sigma^{(i)})^2\Delta t>0$ and
$1 \pm 2\theta\lambda_{(d)} (\sigma^{(i)})^2\Delta t>0$.
Under the quadratic approximation to a portfolio VaR, the optimal alternative distribution
$Q$ minimizing the variance of the importance sampling estimator has $\theta$ satisfying
\begin{equation}\label{key}
\psi'(\theta) = E_{\bar{Q}}[L_b|L_b > r_p],
\end{equation}
where $\psi'(\theta)$ is in (\ref{eq:bootstrap_psiplum}), $E_{\bar{Q}}$ is the expectation under
$P_\theta$, which has the form {\rm (\ref{qdensity})} but replaced by the
three joint distributions: $N(\bar{\mu}_j(\theta),
\bar{\sigma}_j^2(\theta))$, $Pois(\bar{\lambda}(\theta))$, and $\bar{V}(j)$.
\end{corollary}

The optimal $\theta_p$ satisfying (\ref{key}) can be searched by standard numerical methods or the recursive algorithm presented in Section 2.2.
Next we present the importance sampling
algorithm for multi-variate jump diffusion model as follows:\\
\begin{enumerate}
\item
Compute the $\theta_p$ such that
$\psi'(\theta_p)=E_{\theta_p}[L_b|L_b>r_p]$.
\item
\begin{enumerate}
\item[(i)] Generate $(Z_1,\ldots,Z_d)^T \sim (N(\mu_1(\theta_p),\sigma_1^2(\theta_p)),\ldots, N(\mu_d(\theta_p),\sigma_d^2(\theta_p)))^T$.
\item[(ii)] Generate $N(\Delta t)\sim Pois(\lambda(\theta_p))$.
\item[(iii)] Given $N(\Delta t)=n$ generate
\[
V(i)=
(\log Y_i^{(1)},\log Y_i^{(2)},\ldots,\log Y_i^{(d)})^T
\sim N(\eta(\theta_p),
\displaystyle\underline{\delta}^T \Sigma_J\displaystyle\underline{\delta}),~ i=1,\ldots,n.
\]
\end{enumerate}
\item Repeat step 2 $k$ times to have $L_{b,i}$ defined in (\ref{tdiagonal}) for $i=1,\ldots,k$.
Compute $\hat{p}(\theta_p)=\frac{1}{k}\sum_{i=1}^k {\b 1}_{\{L_{b,i} > r_p\}}e^{-\theta_p
(L_{b,i}-r_p)+\psi(\theta_p)}$.
\item Repeat steps 2 and 3 with Monte Carlo size $M$, and compute sample variance $\widehat{var}_{p}
(\theta_p)$.
\end{enumerate}

In Table 4, we compare the relative efficiency (RE) with fifteen risk factors of
the proposed method (\al{PSD}) with respect to the naive method (NV)
in estimating the loss probabilities $P\{L_b > x\}$ with
different values of $x$. Here the relative efficiency,
$\textrm{RE}({\rm Method 1},{\rm Method 2})$, is the variance under
Method 2 divided by that under Method 1. The parameters
are $\lambda=1$, $\eta^{(1)}=\eta^{(2)}=\ldots=\eta^{(15)}=0$,
$\delta^{(i)}=i/100$, $\mu^{(i)}=i/100$, $\sigma^{(i)}=0.1+i/100$
and $\rho_{ij}=0.3$ for all $i\neq j$, sample size $k=1,000$ and
Monte Carlo replication $M=10,000$. In Table 4, we set $b_1=0.044$,
$b_2=0.0589891$, $b_3=0.0720326$, $b_4=0.0838734$, $b_5=0.0948957$,
$b_6=0.105325$, $b_7=0.115304$, $b_8=0.124929$, $b_9=0.134271$,
$b_{10}=0.143378$, $b_{11}=0.152289$, $b_{12}=0.161034$,
$b_{13}=0.169635$, $b_{14}=0.178112$, $b_{15}=0.186479$,
$\lambda_1=5.2$ and $\lambda_2=\ldots=\lambda_{15}=0.7$ with
quadratic approximation function for a fifteen-variate jump
diffusion model. We can see that
the proposed method is significantly more efficient than the
naive Monte Carlo in $10,000$ simulations for all values of $x$.
Furthermore, the efficiency gain is larger for
smaller probabilities against the naive method.

\begin{table}[!ht]\label{T6}
\begin{center}
\caption{Quadratic approximation function compared with
naive method and a multi-variate jump diffusion model in Monte Carlo simulation.}
\vspace{0.3cm}
\begin{tabular}{cccccccccc} \toprule
$r_p$                      &  0.824     & 1.166    &  1.549  \\
$p$                        &  0.0500    & 0.0100   &  0.00100 \\ \midrule
NV $\hat{p}$            &  0.0501    & 0.00998  &  0.00099  \\
$\widehat{var}_p$   &  4.80E-05  & 1.00E-05 &  9.85E-07 \\
PSD $\hat{p}(\theta) $    &  0.0501    & 0.00999  &  0.000100  \\
$\widehat{var}_p(\theta)$   &  5.12E-06  & 2.69E-07 &  3.65E-09 \\
$\textrm{RE}(\widehat{p}(\theta),\widehat{p})$
                           &   9.36  & 37.15 & 269.79 \\ \bottomrule
\end{tabular}
\end{center}
\end{table}
 \noindent {\small $p$ denotes the true tail
probability,  $r_p$ denotes the quantile of $p$, $\hat{p}$ and
$\widehat{var}_p$ are the mean and variance of the probability
estimator with Monte Carlo, $\hat{p}(\theta)$ and
$\widehat{var}_p(\theta)$ are the mean and the variance of the
tail probability estimator with importance sampling,
$\textrm{RE}(\hat{p}(\theta),\hat{p})$ is the relative efficiency
of $\hat{p}(\theta)$ relative to $\hat{p}$ in a multi-variate jump
diffusion model.}

\subsection{Bootstrapping confidence regions in regression model}

Consider a regression model
\begin{eqnarray}\label{reg}
Y_i = \sum_{j=1}^p x_{ij} \beta_j + \varepsilon_i~~~{\rm for}~~~i=1,2,\ldots,n,
\end{eqnarray}
where $p \geq 2$, $\varepsilon_1,\ldots, \varepsilon_n$ are i.i.d.
mean zero random variables with distribution $F(\cdot)$. Denote $\sigma^2$ as the variance of $\varepsilon_1$.
Let ${\bf x_i} = (x_{i1},\ldots,x_{ip})^{T},~\beta=(\beta_1,\ldots,\beta_p)^T$. In vector
and matrix notation, we write
$Y_i =  {\bf x}_{i}^T  {\bf \beta} + \varepsilon_i~~~{\rm for}~~~i=1,2,\ldots,n,$
and
\begin{eqnarray}\label{regm}
{\bf Y} = {\bf X} {\bf \beta} + {\bf \varepsilon},~~~
\end{eqnarray}
where ${\bf X}=(x_{ij})_{n \times p},~{\bf Y}=(Y_1,\ldots,Y_n)^{T}$ and ${\bf \varepsilon}= (\varepsilon_1,\ldots,\varepsilon_n)^{T}$.
Then the least-squares estimator of $\beta$ is
\begin{eqnarray}\label{lse}
\hat{\bf \beta} = ({\bf X}^{T} {\bf X})^{-1} {\bf X}^{T} {\bf Y}.
\end{eqnarray}

Consider the problem of finding a confidence region $C$ in ${\bf R}^p$
that covers the vector $\beta$ with prescribed probability
$(1-\alpha)$. Denote $\beta_0$ as the true parameter. Under the
assumption 
of ${\bf X}$ is full rank $p$, the statistics of interest is
\begin{eqnarray}\label{est}
T := \frac{(\hat{\beta} - \beta_0)^{T}({\bf X}^{T} {\bf X}) (\hat{\beta} - \beta_0)}
{p \hat{\sigma}^{2}},
\end{eqnarray}
where $\hat{\sigma}^{2}$ is the sample variance.
In particular we would like to estimate the probability of the event $\{T\in A\}$,
where $A$ is chosen to be a circular confidence region for the
unknown parameter $\beta_0$. Let $T^*=(\hat{\beta^\ast} -
\hat{\beta})^{T}({\bf X}^{T} {\bf X})(\hat{\beta^\ast} -
\hat{\beta})/p \hat{\sigma^*}^{2}$, where $\hat{\beta^\ast},~\hat{\sigma^*}^{2}$
are the bootstrap estimators of
$\hat{\beta}$ and $\hat{\sigma}^2$, respectively, for given data.
Then the bootstrap estimator of $P\{T\in A\}$
is $\hat{u}=P\{T^*\in A|\hat{\varepsilon}\}=E({\b 1}_{\{T^*\in A\}}|\hat{\varepsilon})$, where
$\hat{\varepsilon}=(\hat{\varepsilon}_1,\ldots,\hat{\varepsilon}_n)$.

To illustrate our proposed method, we select a sample from a regression model in Longley (1967) in
which the data set is available at  \\
http://www.itl.nist.gov/div898/strd/general/dataarchive.html. We choose this
classical data set of labour statistics because it is one of the first used to
test the accuracy of least squares computations. It is noted that the same algorithm can be applied to
current statistical data set.

The response variable ($y$) is the Total Derived Employment and the predictor variables are GNP
Implicit Price Deflator with Year 1954 = 100 ($x_1$), Gross National Product ($x_2$),
Unemployment ($x_3$),
Size of Armed Forces ($x_4$), Non-Institutional Population Age 14 \& Over ($x_5$), and Year ($x_6$).
\begin{eqnarray}\label{long}
y_i=\beta_0+\beta_1 x_{i1}+\beta_2 x_{i2}+\beta_3 x_{i3}+\beta_4 x_{i4}+\beta_5 x_{i5}+\beta_6
x_{i6}+\varepsilon_i~~~i=1,2,\ldots,n
\end{eqnarray}
where the $\varepsilon_i$ is assumed by normal distribution with mean zero and variance
$\sigma^2$ under PMLE.
The least squares estimator (MLE) of $\beta$ is
\[
\hat{\beta}=(\hat\beta_0,\ldots,\hat\beta_6)=(-3482258.63, 15.0618, -0.0358, -2.0202, -1.0332, -0.0511, 1829.15).
\]
The sample variance $\hat{\sigma}^2$ is 55761.60. The bootstrap estimator of $P\{T>a\}$ is $\hat{\alpha}=P\{T^*>a|\hat{\varepsilon}\}=E({\b 1}_{\{T^*>a\}}|\hat{\varepsilon})$,
where $\hat{\varepsilon}=(\hat{\varepsilon}_1,\ldots,\hat{\varepsilon}_n)$.

Let
\[
T^*=\frac{(\hat{\beta}^*-\hat{\beta})^T(X^T X)(\hat{\beta}^*-\hat{\beta})}{p\hat{\sigma^*}^2}=\frac{\sum_{i=1}^n\hat{\varepsilon_i^*}^2}{p\hat{\sigma^*}^2},
\]
where $\hat{\varepsilon_i^*}$ follows normal distribution with mean zero and variance
$\hat{\sigma}^2$.
Here $p=6$ in equation (\ref{long}).

To apply importance resampling in the setting, note that the original bootstrap distribution is
\[
\frac{dP}{d\mathcal{L}}=\prod_{i=1}^n \frac{1}{\sqrt{2\pi} \hat{\sigma}} e^{-\frac{\hat{\varepsilon_i^*}^2}{2\hat{\sigma}^2}}.
\]
The likelihood ratio based on an exponential change of measure is
\begin{eqnarray}\label{normaltilting}
\frac{dQ}{dP}=e^{\theta (T^*-a)-\psi_a(\theta)}=\frac{e^{\theta (T^*-a)}}{Ee^{\theta (T^*-a)}},
\end{eqnarray}
where $\psi_a(\theta)=\log Ee^{\theta (T^*-a)}$.
Let $T_a=T^*-a$.
The moment generating function of $T_a$ is
$ E[e^{\theta T_a}]=e^{-\theta a}(1-2\frac{\theta}{p})^{-\frac{n}{2}}.$
Therefore, the change of measure $Q$ is
\begin{eqnarray*}
\frac{dQ}{d\mathcal{L}}=\frac{dQ}{dP} \frac{dP}{d\mathcal{L}}
= \displaystyle e^{\theta T^*}(1-2\frac{\theta}{p})^{-\frac{n}{2}} \prod_{i=1}^n \frac{1}{\sqrt{2\pi} \hat{\sigma}} e^{-\frac{\hat{\varepsilon_i^*}^2}{2\hat{\sigma}^2}}
= \displaystyle \prod_{i=1}^n \frac{1}{\sqrt{2\pi} \frac{\hat{\sigma}}{\sqrt{1-2\frac{\theta}{p}}}} e^{-\frac{\hat{\varepsilon_i^*}^2}{2\frac{\hat{\sigma}^2}{1-2\frac{\theta}{p}}}}.
\end{eqnarray*}

Recall that $\bar{Q}_\theta = Q_{-\theta}$.
Then, the optimal titling point of our proposed method can be obtained by solving
\begin{equation}\label{tiltingn}
E_{\bar{Q}_\theta} [T^\ast|T^\ast>a]=\psi_a'(\theta)=-a+\log(1-2\frac{\theta}{p})^{-\frac{n}{2}}.
\end{equation}


Instead of doing simulation under the normal family, we propose a transformed likelihood ratio
method to obtain the bootstrap estimator of $P\{T>a\}$.
The method is doing simulation under the $\chi^2$-distribution family.
Note that $T^*={\sum_{i=1}^nX_i^2}/{p}$,
where $X_i={\hat{\varepsilon_i^*}^2}/{\hat{\sigma^*}^2}$ is a $\chi^2$-distribution
with degree of freedom $1$.
Here
the likelihood ratio based on an exponential change of measure is
\begin{eqnarray}\label{chi2tilting}
\frac{dQ}{dP}=e^{\theta (T^*-a)-\psi_a(\theta)}=\frac{e^{\theta (T^*-a)}}{Ee^{\theta (T^*-a)}},
\end{eqnarray}
where $\psi_a(\theta)=\log Ee^{\theta (T^*-a)}$.
By using an argument similar as above, the optimal titling point can be obtained by solving
\begin{equation}\label{tiltingc}
E_{\bar{Q}_\theta} [T^\ast|T^\ast>a]=\psi_a'(\theta)=-a+\log(1-2\frac{\theta}{p})^{-\frac{n}{2}}.
\end{equation}
Note that (\ref{tiltingc}) equals (\ref{tiltingn}).

Since the relative efficiencies of these two algorithms are almost the same, we only report that
based on normal family.
Table 5 reports the relative efficiency of the importance resampling with respect the naive
Monte Carlo simulation. Here $\alpha$ denotes the true tail
probability,  $a$ denotes the quantile of $T^*$, $\hat{\alpha}$ and
$\widehat{var}_\alpha$ are the mean and variance of the probability
estimator with Monte Carlo, $\hat{\alpha}(\theta)$ and
$\widehat{var}_\alpha(\theta)$ are the mean and the variance of the
tail probability estimator with importance sampling,
$\textrm{RE}(\hat{\alpha}(\theta),\hat{\alpha})$ is the relative efficiency
of $\hat{\theta}$ relative to $\hat{\alpha}$ in a regression model.
Table 6 reports the non-coverage probabilities and averages and standard
deviations of region areas for cubical and spherical
confidence regions with nominal coverage probability 95\%.

\begin{table}[!ht]\label{T7}
\begin{center}
\caption{Importance sampling with PMLE compared with naive method and a regression model in
Monte Carlo simulation.}
\vspace{0.3cm}
\begin{tabular}{cccccccccc} \toprule
$a$                                             &  3.363     &  3.756    &  4.571     \\
$\alpha$                                        &  0.100     & 0.050     &  0.010     \\ \midrule
NV $\hat{\alpha}$                            &  0.101       & 0.049   &  0.0099   \\
$\widehat{var}_{\alpha}$                        &  8.94E-04  & 4.74E-04  &  1.00E-04  \\
PSD $\hat{\alpha}(\theta) $                    &  0.099     & 0.050   &  0.010  \\
$\widehat{var}_{\alpha}(\theta)$                &  1.98E-04  & 6.35E-05  &  3.76E-06  \\
$\textrm{RE}(\widehat{\alpha}(\theta),\widehat{\alpha})$    &   4.51  & 13.39 & 26.59 \\ \bottomrule
\end{tabular}
\end{center}
\end{table}

\begin{table}[!ht]\label{T9}
\begin{center}
\caption{ Non-coverage probabilities and averages and standard deviations of region areas for circular confidence regions with nominal coverage probability 95\%.
The sample is drawn from a regression model in Longley (1967). The parameter of interest is $\beta$. Results are reported for studentised statistics and naive resampling
and importance resampling.}
\vspace{0.3cm}
\begin{tabular}{cccccccccc} \toprule
                          & Bootstrap   &              & \multicolumn{2}{c}{Region Area} \\ \cmidrule(l){4-5}
Resampling                & replication & Noncoverage  &               &   Standard  \\
Method                    & size        & probability  &  Average      & deviation   \\  \midrule
Naive                     & 1000        & 0.054        &  49214.4      & 4853.05 \\
Importance                &  400        & 0.049        &  51054.2      & 5035.12 \\
Importance                &  200        & 0.051        &  52563.7      & 5174.63 \\
Importance                &  100        & 0.053        &  54853.3      & 5313.38 \\  \midrule
\end{tabular}
\end{center}
\end{table}

\newpage
\section{Conclusions}

In this paper, we propose a general account in importance sampling
with applications to portfolio VaR computation and bootstrapping confidence regions.
It is shown that our method produces efficient approximation to the problem. Simulation
results confirm the theoretical results that our method always
provides greater variance reduction than the naive Monte Carlo
method. Our numerical experiments demonstrate that the gain in
variance reduction can be substantial in some cases.

The key features of our method are twofold. First, (\ref{eq:theta_ast})
characterizes the optimal alternative distribution for importance
sampling under exponential tilting. And the recursive algorithm facilitates the computation of
the optimal solution. The initial value of the recursive algorithm
is the dominating point of the large deviations tilting probability used previously by other authors;
e.g., Sadowsky and Bucklew (1900). The recursive algorithm then
sequentially generates alternative distributions providing greater
variance reduction. Due to the nature of the recursive algorithm,
the additional programming effort and computing time are negligible.
Second, the proposed tilting formula for normal distribution and its square,
along with jump diffusion model, can be used to have more
efficient simulation for portfolio VaR computation. We also apply the proposed titling probability to
a parametric family (normal family or $\chi^2$ family) which can be used for constructing
bootstrap confidence regions for the unknown parameters in regression models.

Our method highlights the two aforementioned key features in
general settings. By using the idea of conjugate probability measure, we obtained the optimal
tilting parameter via (\ref{eq:theta_ast}). Specific
considerations can be found in several papers; see Fuh and Hu (2004)
for multivariate normal distributions, Fuh and Hu (2007) for hidden
Markov models, and Fuh et al. (2011) for multivariate $t$
distribution. Further applications to $K$-distributions and copula models
will be published elsewhere. In this paper, we assume that
the underlying random variables are independent over time. A more
challenging project is to model the time dependence using, for
example, Markov switching autoregression models or GARCH models.
It is expected that our method can be applied to option pricing, Greeks letters calculation, and correlated
default probabilities among others.\\


\noindent
{\bf Appendix: Proof of Theorem \ref{optimality}}
\def\theequation{A.\arabic{equation}}
\setcounter{equation}{0}\\


The existence of the optimization problem {\rm (\ref{optimization})} can be proved by first showing
that for $\theta \in {\bf R}$,
\begin{eqnarray}\label{id}
 \psi'(\theta) ~{\rm is~ strictly ~increasing~and}~E_{\bar{Q}}[X|X\in A]~{\rm is~strictly~decreasing}.
\end{eqnarray}
To prove (\ref{id}), we first note that $\psi(\theta)$ is the cumulant generating function of $X$,
and therefore its second derivative $\psi''(\theta)>0$ for $\theta \in {\bf I}$. This implies that $\psi'(\theta)$
is strictly increasing. Since $\Psi(\theta)$ is convex and steep by assumption, we have
$\psi'(\theta) \to \infty$ as $\theta \to \theta_{\max}$.
Furthermore, consider the conditional measure of $\bar{Q}$ on the set $A$,
denoted by $\bar{Q}_A$, which is defined as
\begin{equation}\label{eq:tildeQ_A}
d\bar{Q}_A = \frac{\textbf{1}_{\{X\in A \}} d\bar{Q}}{ \int \textbf{1}_{\{X\in A\}} d\bar{Q} }.
\end{equation} Then we have
\begin{eqnarray*}
 \frac{d  E_{\bar{Q}_\theta}[X|X\in A]}{d\theta} &=&
 \frac{d}{d\theta}\left(\frac{E[\textbf{1}_{\{X\in A\}} X  \text{e}^{-\theta X}]}{E[\textbf{1}_{\{X\in A\}}  \text{e}^{-\theta X}]}\right) \\
&=& -\frac{E[\textbf{1}_{\{X\in A\}} X^2  \text{e}^{-\theta X}]} {E[ \textbf{1}_{\{X\in A\}}
 \text{e}^{-\theta X}]} + \frac{E^2 [\textbf{1}_{\{X\in A\}} X  \text{e}^{-\theta X}]}
 {E^2[ \textbf{1}_{\{X\in A\}} \text{e}^{-\theta X}]}
=-{\rm var}_{\bar{Q}_A}(X) < 0.
\end{eqnarray*}
This implies that $E_{\bar{Q}_\theta}[X|X\in A]$ is strictly decreasing. The existence of the optimization problem {\rm (\ref{optimization})}
follows from $E_{\bar{Q}_0}[X|X\in A] = E[X|X\in A] > \mu = \psi'(0)$.

To prove the uniqueness, we
note that the second derivative of $G$ equals
\begin{eqnarray}\label{ddg}
\displaystyle \frac{\partial^2 G(\theta)}{\partial \theta^2}
&=&\frac{\partial^2}{\partial \theta^2} E[\textbf{1}_{\{X\in A\}} \frac{d P}{d Q}]
=\frac{\partial^2}{\partial \theta^2}E[\textbf{1}_{\{X\in A\}}e^{-\theta{X}+\psi(\theta)}] \nonumber \\
&=&\frac{\partial}{\partial \theta} E\al{[\textbf{1}_{\{X\in A\}}
(+\psi'(\theta))e^{-\theta{X}+\psi(\theta)}]} \nonumber \\
&=&E\al{[\textbf{1}_{\{X\in A\}}((-X+\psi'(\theta))^2+\psi''(\theta))e^{-\theta{X}+\psi(\theta)}]}.
\end{eqnarray}
Since $\psi(\theta)$ is the cumulant generating function of $X$, its
second derivative $\psi''(\theta)>0$. It then follows from
(\ref{ddg}) that $\frac{\partial^2 G(\theta)}{\partial \theta^2}>0$,
which implies that there exists a unique minimum of $G(\theta)$.

To prove (\ref{eq:theta_ast}), we need to simplify the RHS of (\ref{abeq}) under $\bar{Q}$. Standard algebra gives
\begin{equation}
\label{eq:ab_temp}
\frac{E[\textbf{1}_{\{X\in A\}}X\mbox{e}^{-\theta X}]}
{E[\textbf{1}_{\{X\in A\}}\mbox{e}^{-\theta X}]}=
\frac{\int \textbf{1}_{\{x\in A\}}x\mbox{e}^{-\theta x} dP / \tilde{\Psi}(\theta)}
{\int \textbf{1}_{\{x\in A\}}\mbox{e}^{-\theta x} dP/ \tilde{\Psi}(\theta)}
= \frac{\int \textbf{1}_{\{x\in A\}}x d\bar{Q}}{\int \textbf{1}_{\{x\in A\}}d\bar{Q}}.
\end{equation}
As a result, (\ref{eq:ab_temp}) equals
$$\int \textbf{1}_{\{x\in A\}} x d\bar{Q}_A = E_{\bar{Q}_A}[X]= E_{\bar{Q}}[X|X\in A],$$
which implies the desired result.

\vspace{1cm}


\centerline{ACKNOWLEDGMENT}

The research of the first author is supported in part by a grant NSC
100-2118-M-008-002-MY3 and NSC 101-3113-P-008-005, the second author is supported in part by a grant
NSC 100-2118-M-008-004, and the third author is supported in part by a grant NSC 100-2410-H-032-024.


\bibliographystyle{nonumber}

\end{document}